\documentclass[final,5p,times,twocolumn]{elsarticle}

\usepackage{amssymb}
\usepackage{amsmath}
\usepackage{lineno}

\journal{Journal of Solid State Chemistry}

\begin{document}

\begin{frontmatter}



\title{Growth, discovery and characterization of single crystalline Eu$_{0.8}$Pt$_6$Al$_{16.4}$}


\author{Juan Schmidt \fnref{label1}}
\fntext[label1]{currently at Instituto de F\'isica de Buenos Aires, CONICET-Universidad de Buenos Aires, Pabell\'on 1, Ciudad Universitaria, CABA, 1428, Argentina.}

\affiliation{organization={Department of Physics and Astronomy, Iowa State University},%Department and Organization
            addressline={}, 
            city={Ames},
            postcode={50011}, 
            state={IA},
            country={USA}}

\affiliation{organization={Ames National Laboratory, Iowa State University},%Department and Organization
            addressline={}, 
            city={Ames},
            postcode={50011}, 
            state={IA},
            country={USA}}

\author{Oliver Janka} %% Author name

\affiliation{organization={Universit{\"a}t des Saarlandes, Anorganische Festk{\"o}rperchemie},%Department and Organization
            addressline={Campus C4 1}, 
            city={Saarbr{\"u}cken},
            postcode={66123}, 
            country={Germany}}

\author{Jutta K\"osters} %% Author name

\affiliation{organization={Institut f{\"u}r Anorganische und Analytische Chemie, Universit{\"a}t M{\"u}nster},%Department and Organization
            addressline={Corrensstrasse 28}, 
            city={M{\"u}nster},
            postcode={48149}, 
            country={Germany}}

\author{Sergey L. Bud'ko} %% Author name

\affiliation{organization={Department of Physics and Astronomy, Iowa State University},%Department and Organization
            addressline={}, 
            city={Ames},
            postcode={50011}, 
            state={IA},
            country={USA}}
            
\affiliation{organization={Ames National Laboratory, Iowa State University},%Department and Organization
            addressline={}, 
            city={Ames},
            postcode={50011}, 
            state={IA},
            country={USA}}

\author{Paul C. Canfield} %% Author name

\affiliation{organization={Department of Physics and Astronomy, Iowa State University},%Department and Organization
            addressline={}, 
            city={Ames},
            postcode={50011}, 
            state={IA},
            country={USA}}

\affiliation{organization={Ames National Laboratory, Iowa State University},%Department and Organization
            addressline={}, 
            city={Ames},
            postcode={50011}, 
            state={IA},
            country={USA}}

%% Abstract
\begin{abstract}
We report the discovery of a ternary compound, Eu$_{0.8}$Pt$_6$Al$_{16.4}$. We determine its chemical and structural characteristics based on energy-dispersive X-ray spectroscopy as well as both powder and single-crystal X-ray diffraction, demonstrating that it crystallizes in a hexagonal structure type EuPt$_6$Al$_{17}$ with no reported structural analog. The temperature- and field-dependent magnetization, and temperature-dependent resistance measurements, reveal that the Eu$^{2+}$ magnetic moments order antiferromagnetically below 2.8 K.
\end{abstract}


%\includegraphics[width=\linewidth]{grabs}


%\item Crystals of Eu$_{0.8}$Pt$_6$Al$_{16.4}$ can be grown by the solution-growth method.
%\item The structural type EuPt$_6$Al$_{15}$ currently has no reported analogs.
%\item One Eu $2e$ site and one Al $6i$ site exhibit vacancies.
%\item Europium exhibits a valence $+2$, and its moments order antiferromagnetically.

%% Keywords

\end{frontmatter}




\section{Introduction} 
\label{sec:Introduction}

An extensive number of compounds with rare-earth, transition metals and aluminum with diverse structures have been found \cite{Morozkin2018}, many of them remain underexplored. In particular, a plethora of rare-earth ($R$) aluminum platinides have been discovered, such as $R$PtAl \cite{Dwight1984}, Ce$_3$Pt$_4$Al$_6$ \cite{Tursina2008}, $R$Pt$_6$Al$_3$ \cite{Eustermann2017}, $R_2$Pt$_9$Al$_{16}$ \cite{Engel2024}, and $R$PtAl$_2$ \cite{Radzieowski2019}, and a broader class of $R$Pt$_{5-x}$Al$_x$ which includes $R$Pt$_3$Al$_2$ as well as $R$Pt$_4$Al \cite{Blazina1995}. Additionally, there have been reports of compounds built of disordered $R$-Al layers and a substantial amount of vacancies in the $R$ sites, such as $R_{1.33}$Pt$_3$Al$_8$ and $R_{0.67}$Pt$_2$Al$_5$, which suggests that a larger cell may be needed to describe these compounds \cite{Latturner2002}. Both of these have related structures made of the same $R$-Al layers but in an ordered arrangement forming the superstructures $R_4$Pt$_9$Al$_{24}$ \cite{Thiede1999} and $R_2$Pt$_6$Al$_{15}$ \cite{Radzieowski2017}, respectively.

Recently, a Eu$_2$Pt$_6$Al$_{15}$ phase was reported to exhibit a valence transition from Eu$^{2+}$ to Eu$^{3+}$ upon cooling below 45 K \cite{Radzieowski2018}, according to M\"ossbauer spectroscopy, resistance and magnetic susceptibility measurements. This phase was obtained in polycrystalline form by arc-melting Eu, Pt and Al together, followed by a thermal treatment. As part of an attempt of obtaining that phase in single crystalline form by solution growth \cite{Schmidt2025arxiv}, we discovered a different ternary: Eu$_{1-x}$Pt$_{6}$Al$_{17-y}$, with $x\sim 0.2$ and $y\sim 0.6$. Since it adopts a structure that has not been reported in any other compound, in this paper we report the details for obtaining single crystals of this phase, as well as its basic structural, chemical, and magnetic characterization.

\section{Experimental details}
\label{sec:Experimental}

\begin{figure*}
\centering
\includegraphics[width=0.9\linewidth]{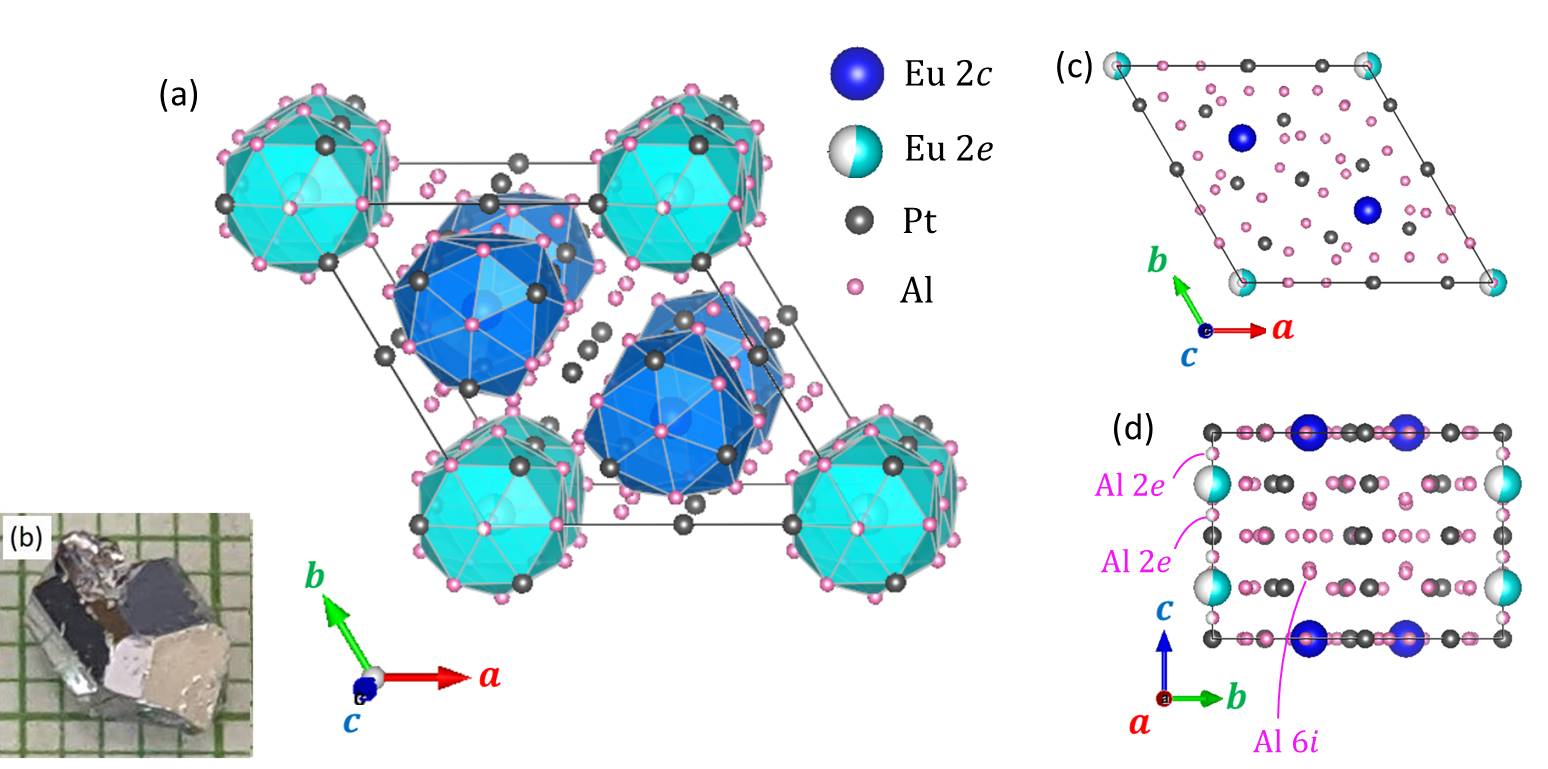}
\caption{{Crystal structure of Eu$_{0.8}$Pt$_6$Al$_{16.4}$, with the Eu atoms in blue/cyan, the Pt atoms in dark gray and the Al atoms in pink: (a) including polyhedra around the Eu atoms in the $2c$ sites without vacancies (blue) and around the Eu atoms in the $2e$ sites with vacancies (cyan). (b) Photograph of one of the obtained single crystals.  (c) Unit cell viewed down the $c$-axis; (d) unit cell viewed down the $a$-axis.}}
\label{fig:EuPtAl_structure}
\end{figure*}

The crystals of Eu$_{0.8}$Pt$_6$Al$_{16.4}$ were grown by a similar method as described in Ref. \citenum{Schmidt2025arxiv}, with a different initial ratio of Eu:Pt:Al of 2:10:88. Platinum powder with purity of 99.9+\%, aluminum rod with purity of 99.999\% and europium pieces with purity of 99.99+\%. The platinum and aluminum were melted together in an arc-furnace \cite{Schmidt2025arxiv} in the ratio 10:88. We placed the resulting button and the missing amount of europium in a 2 ml Canfield Crucible Set \cite{CanfieldP.C.KongT.KaluarachchiU.S.2016}. We used a box furnace to heat the crucible with the elements sealed inside a silica ampule filled with 0.2~atm of argon. It was first kept for 5 hours at 1180~$^{\circ}$C, and then cooled down to 900~$^{\circ}$C over 96 hours. At this temperature, we decanted the solution using a centrifuge \cite{Canfield2020}, obtaining relatively large crystals as the one shown in Fig. \ref{fig:EuPtAl_structure}(b).

The elemental composition of the crystals was determined by energy dispersive X-ray spectroscopy (EDS) using a JOEL scanning-electron microscope with a Thermo NORAN Microanalysis detector. Flat samples of Eu$_2$Pt$_6$Al$_{15}$ \cite{Radzieowski2018,Schmidt2025arxiv} were used as a standard for Eu, Pt and Al quantification. The composition of each crystal was measured at four different positions on arbitrary crystal faces. All the measurements were done with a $16\ \text{kV}$ acceleration voltage, 10~mm working distance, and $35 ^{\circ}$ take-off angle. We used the NIST-DTSA II Microscopium software \cite{Newbury2014} to fit each spectrum, and eventually obtain average compositions and error bars account for the goodness of fit as well as inhomogeneity in the composition. 

X-ray diffraction data were collected on powder samples with a Rigaku MiniFlex II diffractometer with a Cu K$\alpha$ source. The powder was obtained by grinding a few single crystals, and spread over a small layer of vacuum grease on a silicon zero-background holder. The measurements were performed for 2$\theta = 5$–$100^{\circ}$, with a step size of 0.01$^{\circ}$ and a counting time of 4~s per step. The patterns were analyzed by Rietveld refinement in GSAS-II \cite{Dreele2014}.

Single-crystal X-ray diffraction was measured on small regularly shaped crystals. Diffraction frames were collected on a STOE IPDS-II diffractometer equipped with graphite-monochromatized Mo K$\alpha$ radiation and operated in oscillation mode. The small crystals were obtained by fracturing some of the larger as-grown single crystals. The measured intensities were corrected for absorption effects through numerical methods \cite{Stoe2014}.

Magnetization measurements were performed with a Quantum Design Magnetic Property Measurement System (MPMS Classic and MPMS3) device based on a superconducting quantum interference magnetometers, covering 1.8~K~$\leq T \leq$~300~K and fields up to $|H| = 70$~kOe. Each crystal was oriented by Laue diffraction so that the magnetic field could be applied along specific crystallographic directions. Both zero-field cooled (ZFC) and field-cooled (FC) protocols were used for the measurements. Samples were fixed to a Kel-F disk mounted inside a plastic straw; the magnetic signal of the mounting disk was measured separately and subtracted from the total moment.

 We used the AC Transport option of a Quantum Design Physical Property Measurement System (PPMS) to measure the resistance of the samples as a function of temperature. A current of 3~mA and 17~Hz was applied both parallel and perpendicular to the c-axis, employing a standard four-probe configuration. By spot-welding 25~$\mu$m platinum wires we were able to achieve electrical contacts with resistances below 1.5 $\Omega$. Additionally we applied Epo-Tek H20E silver epoxy and cured it at 120$^{\circ}$C for 1 h, to improve the mechanical robustness of the contacts.

\section{Results and discussions}
\label{sec:Results}

\subsection{Chemical and structural characterization}

\begin{table}[h!tb]
\centering
\caption{\label{tab:scxrd_EuPtAl}Single crystal X-ray diffraction measurement and refinement information for Eu$_{0.8}$Pt$_6$Al$_{16.4}$.}
\begin{tabular}{ll}
\hline
\hline
\textrm{Chemical formula}&
\textrm{Eu$_{0.781(5)}$Pt$_6$Al$_{16.4(1)}$}\\
\hline
Formula Weight & 1730.2 \\
Temperature & 293 K \\
Crystal system & hexagonal\\
Space group & $P\bar{6}2m$\\
 & $a=14.2254(6)\ \text{\AA}$ \\
Unit cell dimensions & $c=8.7363(4)\ \text{\AA}$ \\
Volume & 1531.04(12) $\text{\AA}^3$ \\
$Z$ & 4 \\
Calculated density & 7.51 g/cm$^3$\\
Absorption coefficient & 58.7 mm$^{-1}$\\
$F(000)$ & 2894\\
Radiation & Mo K$\alpha$ ($\lambda=0.71073\ \text{\AA}$)\\
$2\Theta$ range for collection & 4.66$^{\circ}$ to 66.81$^{\circ}$\\
Min/Max index [$h$, $k$, $l$] & [$-$21/20, $-$21/20, $-$13/13]\\
Reflections collected & 18620\\
Independent reflections & 2204 [$R_{int}=0.0543$]\\
Data/restraint/parameters & 2204/1/100\\
Goodness-of-fit on $F^2$ & 1.25\\
Final $R$ indexes [$I\geq 2\sigma (I)$] & $R_1=0.0207$, $wR_2=0.0208$\\
Final $R$ indexes [all data] & $R_1=0.0387$, $wR_2=0.02245$\\
Largest diff. peak/hole & 2.71 $e^- \text{\AA}^{-3}$/$-$2.26 $e^- \text{\AA}^{-3}$\\
\hline
\hline
\end{tabular}
\end{table}

Single-crystal X-ray diffraction measurement on the obtained crystals of Eu$_{0.8}$Pt$_6$Al$_{16.4}$, revealed that it adopts a crystal structure Eu$_{1-x}$Pt$_6$Al$_{17-y}$ ($x\sim 0.2$, $y\sim 0.6$) that does not match with any currently reported one. Table \ref{tab:scxrd_EuPtAl} shows the information on the measurement and refinement. The system is hexagonal and belongs to the space group $P\bar{6}2m$. A diagram of the unit cell basis of the structure is presented in Fig. \ref{fig:EuPtAl_structure}, and the detailed information of the atomic positions, occupancies and thermal displacement parameters are given in Tab. \ref{tab:scxrd_EuPtAl_2}. There are two distinct Eu sites ($2c$ and $2e$) located in the center of the blue and cyan polyhedra defined by the surrounding Pt and Al atoms in Fig. \ref{fig:EuPtAl_structure}(a). The Eu $2e$ sites (in the cyan polyhedra) have an occupancy of around 56(1)\%, whereas the other has full occupancy. Besides the Eu $2e$ position, also the Al $6i$ and Al $2e$ positions show an occupancy lower than 1, while all other positions are fully occupied within three standard deviations. The two Al $2e$ atoms form Al$_2$ dumbbells that occupy the same space as the Eu $2e$ atoms, therefore only the Eu atom or the Al dumbbell can be present at the same time. This is also reflected by the occupancy factors of the Eu $2e$ (56\%) and the Al dumbbell (46\%), both obtained in a free refinement. The occupancy of the Al $2e$ positions was constrained to be equal. The resulting composition is therefore Eu$_{0.781(5)}$Pt$_6$Al$_{16.4(1)}$, corresponding to a $x=0.219(5)$ and a $y=0.6(1)$. Since the Eu occupation is close to 1/2, different symmetry reductions (subgroup $P3_1m$ of index 2 and subgroup $Cm2m$ of index 3 alongside the introduction of the repective twins/trillings) were tested, however, no conclusive structural model was obtained in a lower symmetric space group. When looking at the structure refinement and the site occupancies, an ordering might occur when changing the composition so that both the Eu $2e$ position and the Al dumbbell are occupied by 50\%. This might lead to a doubling of the $c$-axis allowing an ordering of dumbbells and Eu atoms.

%A potential symmetry loss, however, might be indicative by the remaining residual electron density.

\begin{table}[h!tb]
\centering
\caption{\label{tab:scxrd_EuPtAl_2} Refined atomic positions, site occupancy, and isotropic thermal displacement parameters for Eu$_{0.8}$Pt$_6$Al$_{16.4}$, according to SCXRD.}
\resizebox{0.48\textwidth}{!}
{\begin{tabular}{lcccccc}
\hline
\hline
At. & Site & $x$ & $y$ & $z$ & $U_{\text
{iso}}$ & Occ. \\
\hline
Eu & $2c$ & 1/3 & 2/3 & 0 & 0.0085(4) & 1\\
Eu & $2e$ & 0 & 0 & 0.2484(4) & 0.0073(8) & 0.56(1)\\
Pt & $3f$ & 0.5260(1) & 0 & 0 & 0.0059(4) & 1 \\
Pt & $3f$ & 0.8193(1) & 0 & 0 & 0.0065(6) & 1\\
Pt & $3g$ & 0.5183(1) & 0 & 1/2 & 0.0081(4) & 1\\
Pt & $3g$ & 0.8186(1) & 0 & 1/2 & 0.0059(6) & 1\\
Pt & $12l$ & 0.20935(4) & 0.45852(4) & 0.24829(8) & 0.0057(2) & 1\\
Al & $2e$ & 0 & 0 & 0.398(2) & 0.011(6) & 0.46(4) \\
Al & $2e$ & 0 & 0 & 0.101(4) & 0.016(8) & 0.46(4) \\
Al & $3f$ & 0.186(1) & 0 & 0 & 0.011(4) & 1 \\
Al & $3g$ & 0.179(1) & 0 & 1/2 & 0.010(4) & 1 \\
Al & $4h$ & 1/3 & 2/3 & 0.3427(8) & 0.008(2) & 1 \\
Al & $6i$ & 0.3352(6) & 0 & 0.3196(4) & 0.017(2) & 0.93(4) \\
Al & $6j$ & 0.1066(6) & 0.4285(6) & 0 & 0.007(2) & 1 \\
Al & $6j$ & 0.2737(6) & 0.3918(6) & 0 & 0.007(2) & 1 \\
Al & $6k$ & 0.2753(6) & 0.3919(8) & 1/2 & 0.010(4) & 1 \\
Al & $6k$ & 0.1632(8) & 0.498(1) & 0.5 & 0.0163(4) & 1 \\
Al & $12l$ & 0.1158(4) & 0.5850(4) & 0.2432(8) & 0.007(1) & 1 \\
Al & $12l$ & 0.1432(4) & 0.2567(4) & 0.2502(8) & 0.012(2) & 1 \\
\hline
\hline
\end{tabular}}
\end{table}
\medskip

The atomic percentages of Eu, Pt and Al were also quantified by EDS and are compared with the atomic percentages given by the SCXRD in Tab. \ref{tab:EDS_EuPtAl}. The two results are in agreement within the uncertainties of each analysis, stated in the Tab. \ref{tab:EDS_EuPtAl} as 95\% confidence intervals.

\begin{figure}[h!tb]
\centering
\includegraphics[width=\linewidth]{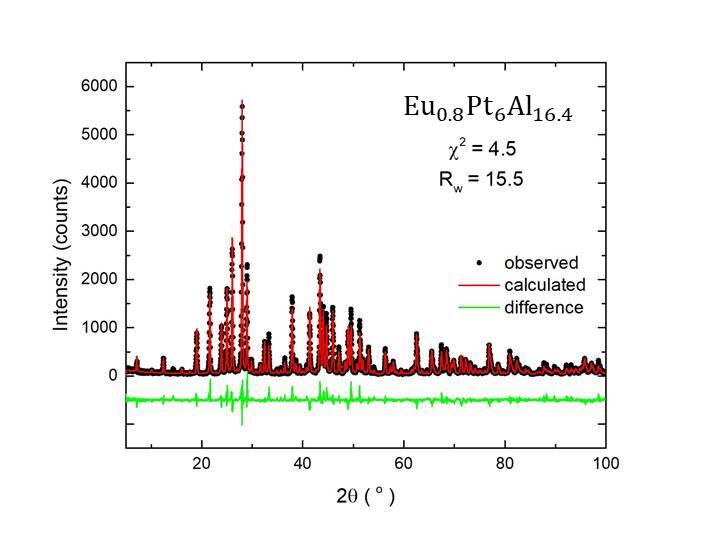}
\caption{\footnotesize{Powder X-ray diffraction of Eu$_{0.8}$Pt$_6$Al$_{16.4}$ showing measured intensities (black points), the Rietveld-refined profile (red trace), and difference curve (green).}}
\label{fig:powder_EuPtAl}
\end{figure}

Some of the crystals obtained were ground in order to measure powder X-ray diffraction, obtaining the pattern shown in Fig. \ref{fig:powder_EuPtAl}, plotted with black symbols. A Rietveld refinement was performed, yielding $\chi^2=4.5$ and $R_w=15.5$. The model structure obtained from the single crystal X-ray refinement was used as the starting point; only the global factor, lattice parameters, atomic positions, microstrains, sample displacement and background were refined. The rest of the parameters, including the thermal displacements and the atomic fractions were kept fixed according to the values reported for the single crystal. The Rietveld-refined profile is plotted as a red curve, showing good agreement with the measured data, as also indicated by the difference (green line).

\begin{table}[h!tb]
\centering
\caption{\label{tab:EDS_EuPtAl} Elemental analysis of Eu$_{0.8}$Pt$_6$Al$_{16.4}$.}
\begin{tabular}{ccc}
\hline
\hline
\ \ \ \textrm{Element}\ \ \ &\ \ \ 
\textrm{from EDS}\ \ \ &\ \ \ \textrm{from SCXRD}\ \ \ \\
\hline
Eu & 3.6(2) & 3.37(4) \\
Pt & 28(2) & 25.9(1) \\
Al & 68(2) & 70.7(8)\\
\hline
\hline
\end{tabular}
\end{table}

\subsection{Electronic and magnetic characterization}

\begin{figure}[h!tb]
\centering
\includegraphics[width=\linewidth]{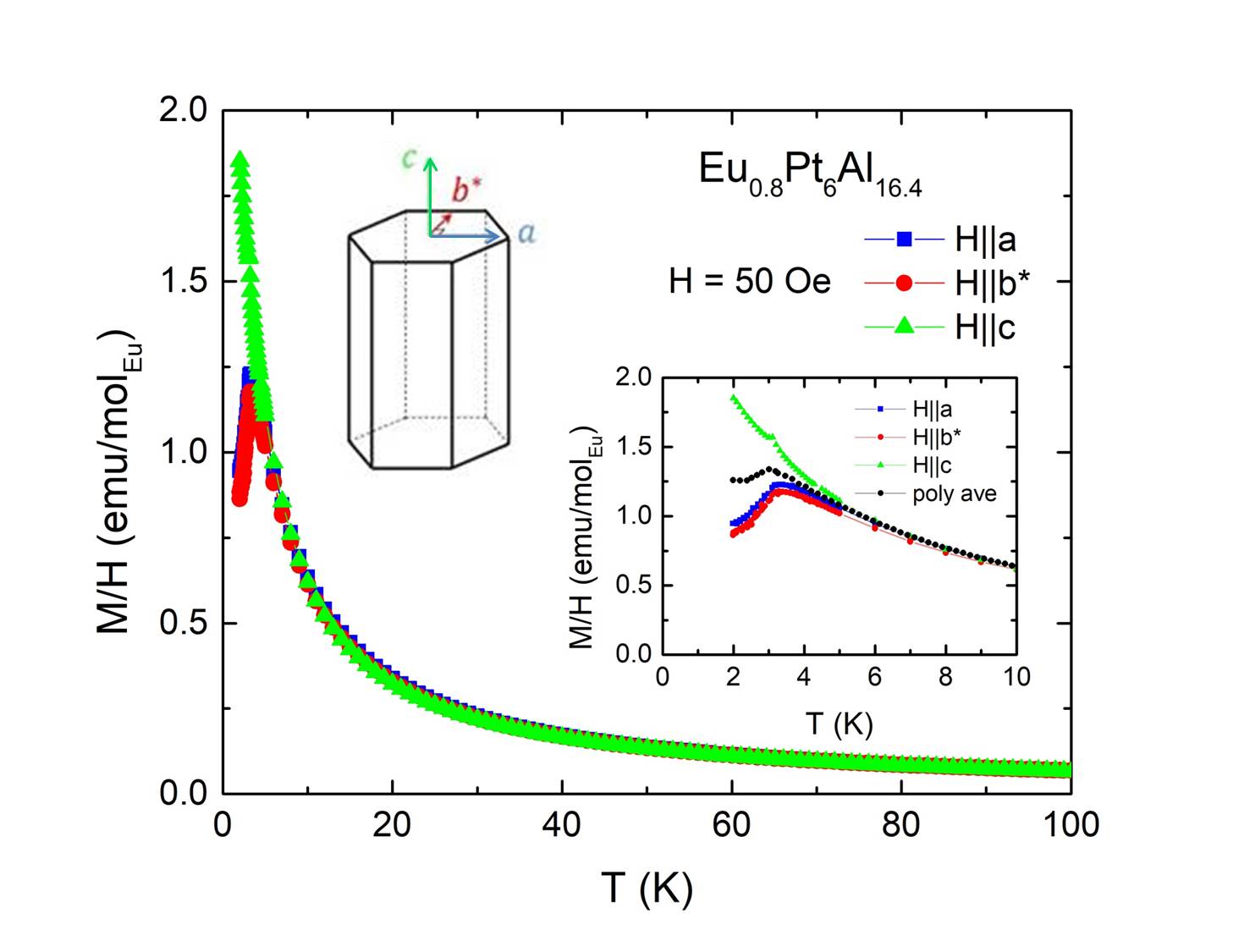}
\caption{\footnotesize{Magnetization as a function of temperature for Eu$_{0.8}$Pt$_6$Al$_{16.4}$ measured with 50 Oe along the $a$, $b^*$, and $c$ axes (blue, red and green, respectively). Data taken under zero-field- and field-cooled conditions overlap throughout the range. The inset expands the low-temperature region to emphasize the observed feature and includes the polycrystalline average (black).}}
\label{fig:MT_massive}
\end{figure}

Fig. \ref{fig:MT_massive} shows the ZFC and FC temperature-dependent magnetization normalized by the applied field of 50 Oe, applied along three mutually orthogonal directions: along the $a$-axis (in blue), along the $c$-axis (in green), and along the reciprocal lattice vector $b^*$ (in red). These three directions are schematically indicated with arrows on a prism drawing in the main panel. The inset shows an enlarged view of the low temperature range of these results, in order to see the feature more clearly in all three directions of applied field. The magnetization along $a$ and $b^*$ shows a downturn upon cooling below 3 K, whereas the magnetization along $c$ shows a more subtle kink and remains higher at low temperatures. The ZFC and FC data remain indistinguishable from each other throughout the measured temperature range. 

\begin{figure}[h!tb]
\centering
\includegraphics[width=\linewidth]{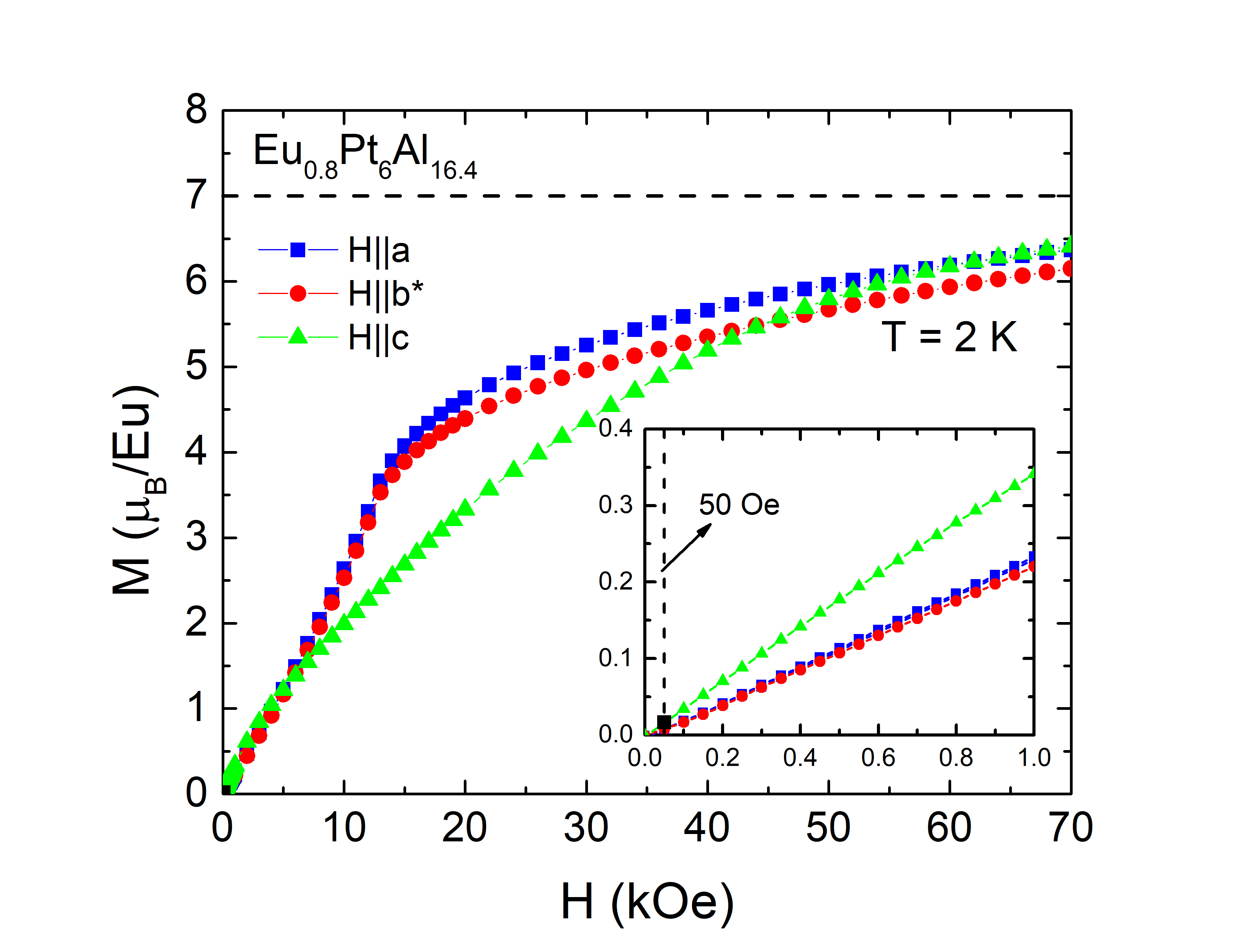}
\caption{\footnotesize{Field-dependent magnetization at 2 K for Eu$_{0.8}$Pt$_6$Al$_{16.4}$ with field applied along $a$ (blue), $b^*$ (red), and $c$ (green). The dashed horizontal line marks the moment expected for Eu$^{2+}$ ($J = 7/2$). Inset: low-field segment showing linear behavior and the field used for the temperature scans. The overlapping black squares show the corresponding ZFC and FC magnetization at 2 K obtained from the $M(T)$ measurements done with a field of 50 Oe applied parallel to $c$.}}
\label{fig:MH_massive}
\end{figure}

The magnetization, in units of Bohr magneton, $\mu_B$, per Eu atom is plotted as a function of the applied field along $a$, $b^*$ and $c$ in Fig. \ref{fig:MH_massive}, measured at a constant temperature of 2 K with an increasing magnetic field. As was the case for Fig. \ref{fig:MT_massive}, the formula unit of Eu$_{0.781(5)}$Pt$_6$Al$_{16.4(1)}$ was used for computing the intrinsic value of magnetization per Eu. The inset shows an enlarged view of the low-field data, showing a linear behavior that extrapolates to $M(H=0)=0$. No hysteresis is expected in any of the three directions, since no difference was observed between the temperature-dependent ZFC and FC measurements done at 50 Oe (see Fig. \ref{fig:MT_massive}). As a reference, the black squares in the inset of Fig. \ref{fig:MH_massive} show the corresponding ZFC and FC magnetization values extracted from the $M(T)$ measurements at 2 K and a field of 50 Oe applied parallel to $c$, both consistent with the $M(H)$ results. The measurements along $a$ and $b^*$ both display some metamagnetic transition around 12 kOe, which is broadened due to the fact that the measurements were taken at a temperature that is close to the transition temperature. Combined, the $M(T)$ and $M(H)$ measurements shown in Figs. \ref{fig:MT_massive} and \ref{fig:MH_massive} indicate that the sample at low temperatures exhibits antiferromagnetic ordering, without signs of any ferromagnetic component. 

The magnetization shown in Fig. \ref{fig:MH_massive} does not reach full saturation at the maximum field measured, but it is consistent with a saturation of 7 $\mu_B$ expected for Eu$^{2+}$ ions with $J=7/2$, indicated with a horizontal dashed line in the figure when using the proper stoichiometry of Eu$_{0.781(5)}$Pt$_6$Al$_{16.4(1)}$. The $\sim 5\%$ difference between the magnetization for $H||a$ (blue curve) and $H||b^*$ (red curve) could be due to slight displacements of the sample with respect to the central axis of the MPMS. 

\begin{figure}[h!tb]
\centering
\includegraphics[width=\linewidth]{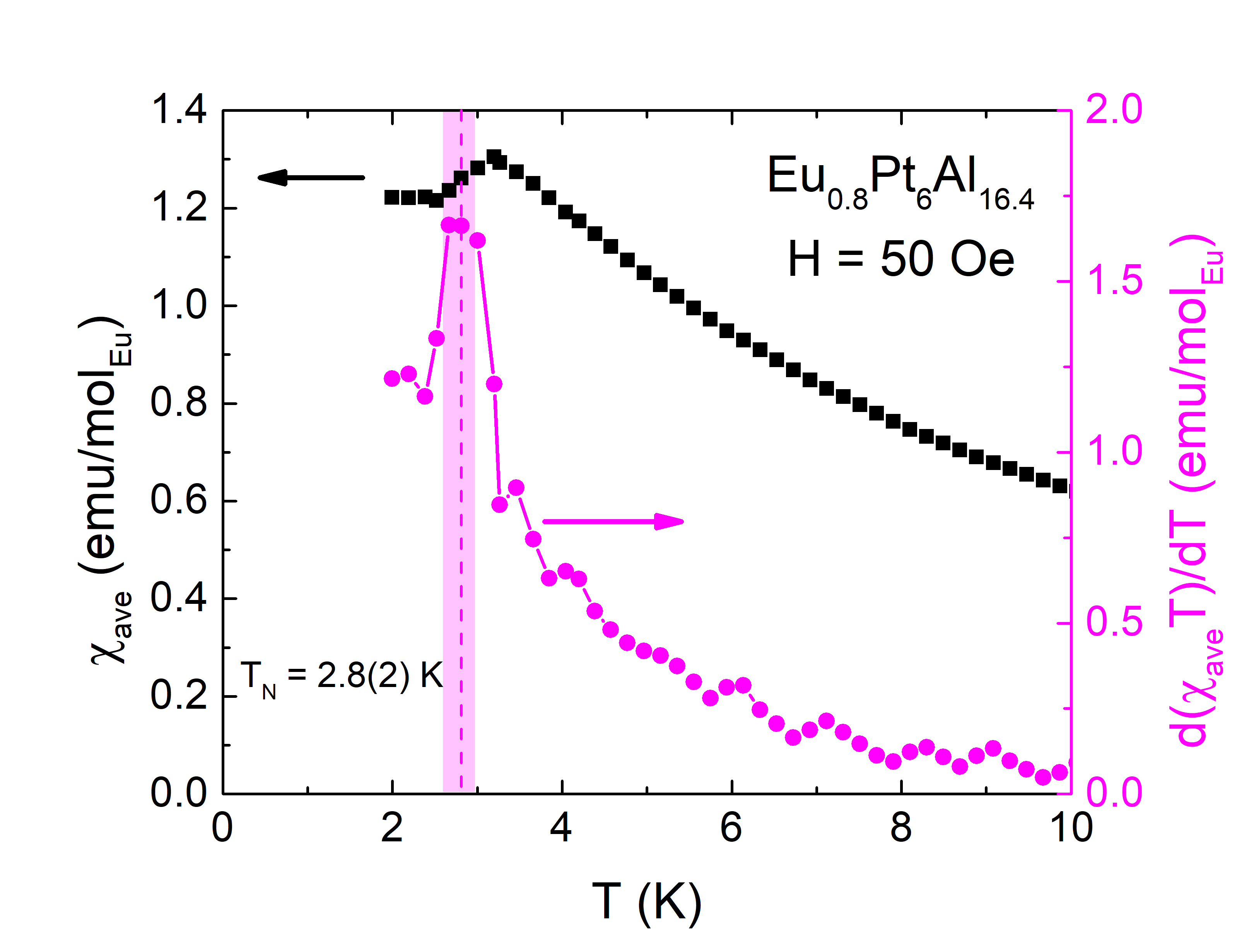}
\caption{\footnotesize{Temperature evolution of the polycrystalline-averaged magnetic susceptibility $\chi_{\text{ave}}$ (black) and its derivative $d(\chi_{\text{ave}}T)/dT$ (magenta). The vertical dashed line identifies the Néel temperature $T_N = 2.8(2)$ K obtained from the derivative maximum.}}
\label{fig:dchiTdT_massive}
\end{figure}

Figure \ref{fig:dchiTdT_massive} shows the polycrystalline average of the three directions (in black) computed as 
\begin{equation}
    \chi_{\text{ave}}=\frac{M_a+M_{b^*}+M_c}{3H},
    \label{eq:polycrystalline}
\end{equation}
for $H=50$ Oe which is within the linear regime of the field-dependent magnetization along the three directions. To obtain an accurate value of the transition temperature, $d(\chi_{\text{ave}} T)/dT$ was computed, since this quantity is expected to be proportional to the specific heat for temperatures close to a second-order antiferromagnetic transition \cite{Fisher1962}. The temperature dependence of $d(\chi_{\text{ave}} T)/dT$ is plotted in magenta in Fig. \ref{fig:dchiTdT_massive}, and the N\'eel temperature, $T_N=2.8(2)$ K, is determined from its maximum, as indicated with the vertical dashed line. 

\begin{figure}[h!tb]
\centering
\includegraphics[width=\linewidth]{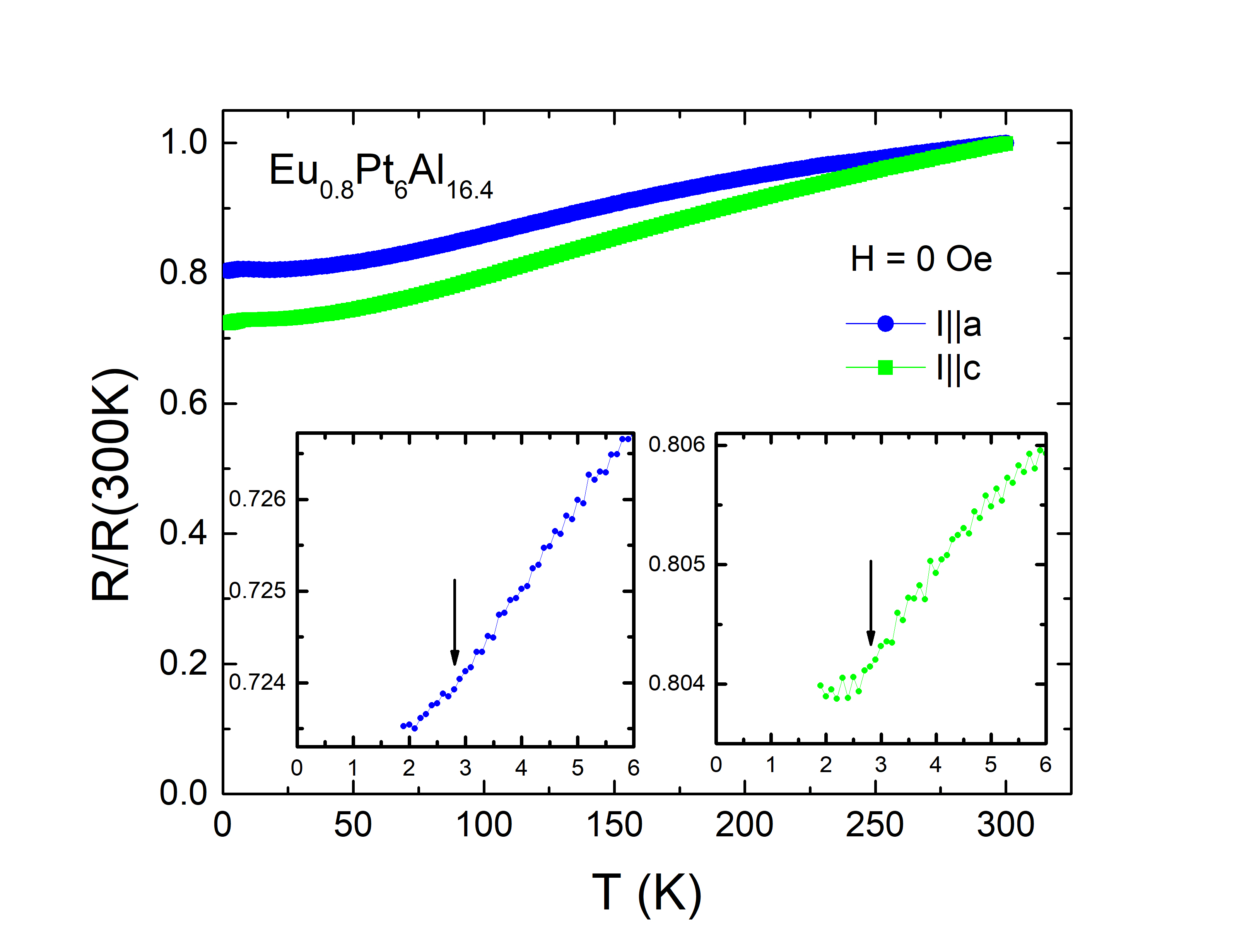}
\caption{\footnotesize{Resistivity normalized to the room-temperature value for Eu$_{0.8}$Pt$_6$Al$_{16.4}$, measured with current along $a$ (blue) and $c$ (green) during cooling and warming cycles. Insets enlarge the low-temperature region; the black arrows indicate the estimated transition temperature according to $d(\chi_{\text{ave}}T)/dT$.}}
\label{fig:RT_massive}
\end{figure}

Fig. \ref{fig:RT_massive} shows the resistance, normalized to the room temperature value, as a function of temperature, for the current applied along $a$ (blue) and along $c$ (green). The residual resistivity ratio (RRR) is less than 1.5 for both directions, indicating a predominance of temperature-independent defect scattering among other types of scattering, most likely as a consequence of a highly disordered structure. The insets show an enlarged view of the low-temperature range for each direction, and the black arrows indicate the transition temperature according to the magnetic susceptibility measurements. It can be noted that the resistance does not show any discernible feature at the temperature corresponding to the transition in the magnetic susceptibility.

\begin{figure}[h!tb]
\centering
\includegraphics[width=\linewidth]{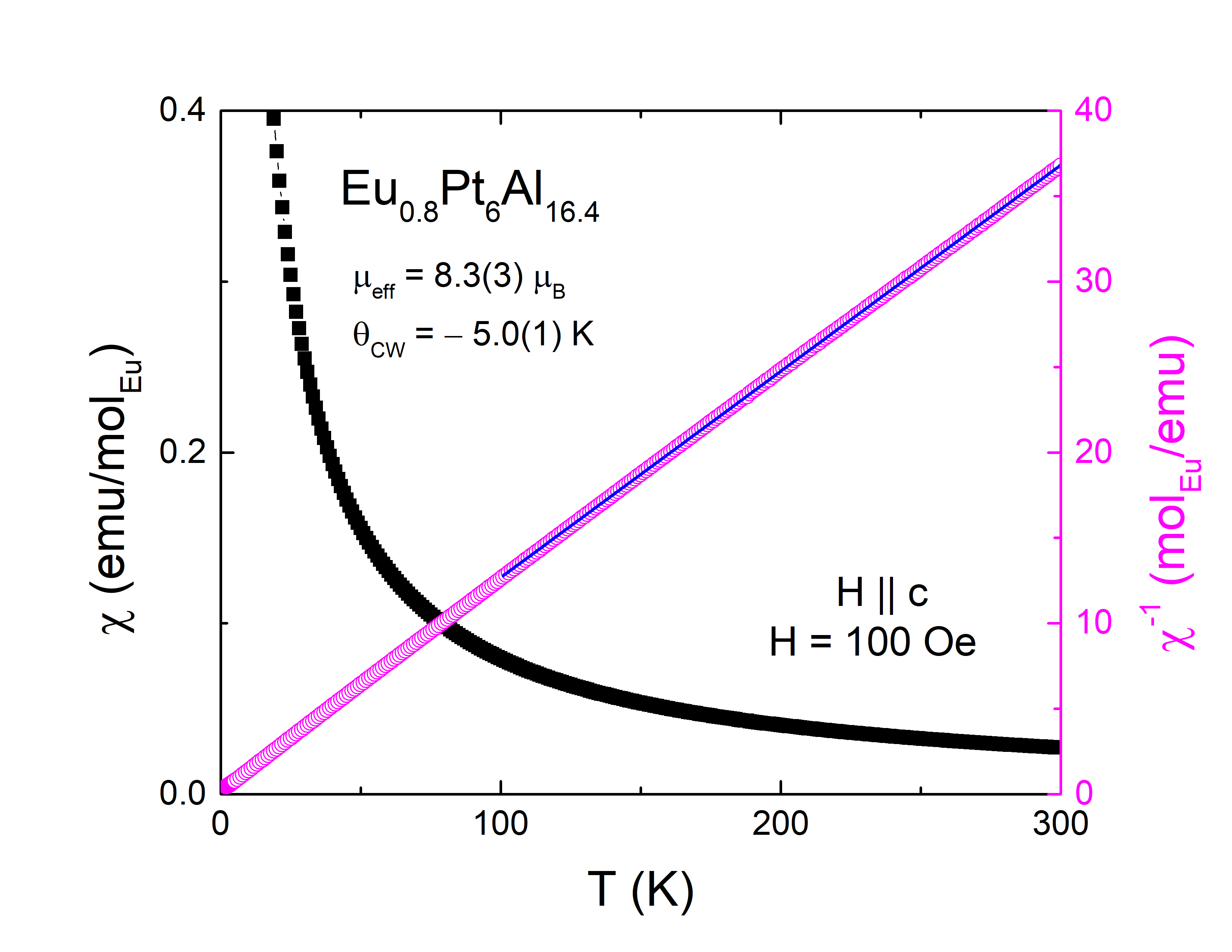}
\caption{\footnotesize{Magnetic susceptibility for Eu$_{0.8}$Pt$_6$Al$_{16.4}$ measured with a 100 Oe field along $c$ (solid symbols, left axis) and its reciprocal (open symbols, right axis). The straight line represents a Curie–Weiss fit yielding $\mu_{eff} = 8.3(3) \mu_B$ and $\theta_{CW} = –5.0(1)$~K, consistent with Eu$^{2+}$ behavior.}}
\label{fig:CW_massive}
\end{figure} 

Magnetic susceptibility was measured up to 300 K applying a field of 100 Oe in order to improve the signal at high temperatures, which decreases asymptotically, and still ensure to lie within the linear regime of $M(H)$, as shown in the inset of Fig. \ref{fig:MH_massive} even in the ordered state. It was only measured with the field applied parallel to $c$, since the susceptibility is isotropic in the paramagnetic state, as expected for Eu$^{2+}$ moments with $L=0$, and as shown in the main panel of Fig. \ref{fig:MT_massive}. This susceptibility and its inverse are plotted in Fig. \ref{fig:CW_massive} in black filled symbols (left axis) and magenta open symbols (right axis), respectively. The latter displays a linear dependence as expected for a Curie-Weiss law behavior. A linear fit was done for $100 \leq T \leq 300$ K in order to obtain the effective moment, $\mu_{\text{eff}}$ and the Weiss temperature, $\theta_{CW}$, according to
\begin{equation}
    \chi^{-1}=\frac{\mu_{\text{eff}}^2 N_A}{3k_B}(T-\theta_{CW})
    \label{eq:CW1}
\end{equation}
This yielded $\mu_{\text{eff}}=8.3(3)\ \mu_B$, which agrees with the expected value for Eu$^{2+}$ ions (7.94 $\mu_B$) within uncertainty; and $\theta_{CW}=-5.0(1)$ K, which would indicate the paramagnetic phase has predominantly antiferromagnetic interactions, and is consistent with the low-temperature antiferromagnetic ordering.

\section{Conclusion}
\label{sec:Conclusion}

A ternary compound of Eu, Pt and Al with no reported structural analogs was discovered and characterized. It was obtained in single crystalline form out of a ternary melt with excess Al.

Its chemical formula can be expressed as Eu$_{1-x}$Pt$_{6}$Al$_{17-y}$, with $x=0.219(5)$ and $y=0.6(1)$, as determined by refining the occupancy of the single crystal X-ray diffraction data. The chemical composition determined from energy dispersive X-ray spectroscopy was consistent. The structure is hexagonal and belongs to the $P\bar{6}2m$ space group. The characterization measurements done on this phase are consistent with the Eu$^{2+}$ moments ordering antiferromagnetically below 2.8(2) K.

\section{Acknowledgements}
J.S. thanks Dolores Mar\'ia Bolo for helping with the graphics design. Work at Ames National Laboratory (J.S., S.L.B., and P.C.C.) was supported by the U.S. Department of Energy, Office of Basic Energy Science, Division of Materials Sciences and Engineering. Ames National Laboratory is operated for the U.S. Department of Energy by Iowa State University under Contract No. DE-AC02-07CH11358. Work at Universitat des Saarlandes (O.J.) was supported by the German Research Foundation DFG (JA 1891-10-1).

\section{Data Availability}
The data that support the findings of this article are openly
available \cite{data_EuPt6Al16}. The .cif file obtained from the single crystal x-ray diffraction measurement and refinement is available as Supplemental Material \cite{SI}.


\begin{thebibliography}{10}
\expandafter\ifx\csname url\endcsname\relax
  \def\url#1{\texttt{#1}}\fi
\expandafter\ifx\csname urlprefix\endcsname\relax\def\urlprefix{URL }\fi
\expandafter\ifx\csname href\endcsname\relax
  \def\href#1#2{#2} \def\path#1{#1}\fi

\bibitem{Morozkin2018}
A.~Morozkin, A.~Garshev, A.~Knotko, V.~Yapaskurt, Y.~Mozharivskyj, F.~Yuan, J.~Yao, R.~Nirmala, S.~Quezado, S.~Malik, {The Gd-Co-Al system at 870/1070 K as a representative of the rare earth-Co-Al family and new rare-earth cobalt aluminides: Crystal structure and magnetic properties}, J. Solid State Chem. 261 (2018) 62--74.
\newblock \href {https://doi.org/https://doi.org/10.1016/j.jssc.2018.02.009} {\path{doi:https://doi.org/10.1016/j.jssc.2018.02.009}}.

\bibitem{Dwight1984}
A.~Dwight, {Crystal structure of equiatomic ternary compounds: lanthanide-transition metal aluminides}, J. Less-Common Met. 102~(1) (1984) L9--L13.
\newblock \href {https://doi.org/https://doi.org/10.1016/0022-5088(84)90401-6} {\path{doi:https://doi.org/10.1016/0022-5088(84)90401-6}}.

\bibitem{Tursina2008}
A.~I. Tursina, A.~V. Gribanov, N.~G. Bukhan’ko, P.~Rogl, Y.~D. Seropegin, \href{https://doi.org/10.30970/cma1.0027}{{Crystal structure of the novel compound Ce$_3$Pt$_4$Al$_6$}}, Chem. Met. Alloys~(1,№ 1) (2008) 62--66.
\newline\urlprefix\url{https://doi.org/10.30970/cma1.0027}

\bibitem{Eustermann2017}
F.~Eustermann, F.~Stegemann, K.~Renner, O.~Janka, {Platinum Triangles in the Pt/Al Framework of the Intermetallic $RE$Pt$_6$Al$_3$ ($RE =$ Ce–Nd, Sm, Gd, Tb) Series}, Z. Anorg. Allg. Chem. 643~(23) (2017) 1836--1843.
\newblock \href {https://doi.org/https://doi.org/10.1002/zaac.201700090} {\path{doi:https://doi.org/10.1002/zaac.201700090}}.

\bibitem{Engel2024}
S.~Engel, L.~Schumacher, O.~Janka, {Modifying the valence phase transition in Eu$_2$Al$_{15}$Pt$_6$ by the solid solutions Eu$_2$Al$_{15}$(Pt$_{1-x}T _x$ )$_6$ ($T =$ Pd, Ir, Au; $x =$ 1/6)}, Z. Naturforsch. B 79~(1) (2024) 21--27.
\newblock \href {https://doi.org/https://doi.org/10.1515/znb-2023-0072} {\path{doi:https://doi.org/10.1515/znb-2023-0072}}.

\bibitem{Radzieowski2019}
M.~Radzieowski, F.~Stegemann, C.~Doerenkamp, S.~F. Matar, H.~Eckert, C.~Dosche, G.~Wittstock, O.~Janka, {Correlations of Crystal and Electronic Structure via NMR and X-ray Photoelectron Spectroscopies in the $RETM$Al$_2$ ($RE =$ Sc, Y, La–Nd, Sm, Gd–Tm, Lu; $TM =$ Ni, Pd, Pt) Series}, Inorg. Chem. 58~(10) (2019) 7010--7025.
\newblock \href {https://doi.org/https://doi.org/10.1021/acs.inorgchem.9b00648} {\path{doi:https://doi.org/10.1021/acs.inorgchem.9b00648}}.

\bibitem{Blazina1995}
Z.~Blazina, {Structural and magnetic studies of the $RE$Pt$_{5-x}$M$_x$ ($RE=$La, Nd; $M=$Al, Ga, In) systems}, J. Alloys Compd. 216~(2) (1995) 251--254.
\newblock \href {https://doi.org/https://doi.org/10.1016/0925-8388(94)01279-Q} {\path{doi:https://doi.org/10.1016/0925-8388(94)01279-Q}}.

\bibitem{Latturner2002}
S.~E. Latturner, M.~G. Kanatzidis, {Gd$_{1.33}$Pt$_3$(Al,Si)$_8$ and Gd$_{0.67}$Pt$_2$(Al,Si)$_5$: Two Structures Containing a Disordered Gd/Al Layer Grown in Liquid Aluminum}, Inorg. Chem. 41~(21) (2002) 5479--5486.
\newblock \href {https://doi.org/https://doi.org/10.1021/ic025623n} {\path{doi:https://doi.org/10.1021/ic025623n}}.

\bibitem{Thiede1999}
V.~M.~T. Thiede, B.~Fehrmann, W.~Jeitschko, {Ternary Rare Earth Metal Palladium and Platinum Aluminides $R_4$Pd$_9$Al$_{24}$ and $R_4$Pt$_9$Al$_{24}$}, Z. Anorg. Allg. Chem. 625~(9) (1999) 1417--1425.
\newblock \href {https://doi.org/https://doi.org/10.1002/(SICI)1521-3749(199909)625:9<1417::AID-ZAAC1417>3.0.CO;2-S} {\path{doi:https://doi.org/10.1002/(SICI)1521-3749(199909)625:9<1417::AID-ZAAC1417>3.0.CO;2-S}}.

\bibitem{Radzieowski2017}
M.~Radzieowski, F.~Stegemann, R.-D. Hoffmann, O.~Janka, {The monoclinic superstructure of the $M_2$Pt$_6$Al$_{15}$ series ($M=$ Ca, Sc, Y, La, Lu)}, Z. Kristallogr. Cryst. Mater. 232~(10) (2017) 675--687.
\newblock \href {https://doi.org/https://doi.org/10.1515/zkri-2017-2050} {\path{doi:https://doi.org/10.1515/zkri-2017-2050}}.

\bibitem{Radzieowski2018}
M.~Radzieowski, F.~Stegemann, T.~Block, J.~Stahl, D.~Johrendt, O.~Janka, {Abrupt Europium Valence Change in Eu$_2$Pt$_6$Al$_{15}$ around 45 K}, J. Am. Chem. Soc. 140~(28) (2018) 8950--8957.
\newblock \href {https://doi.org/https://doi.org/10.1021/jacs.8b05188} {\path{doi:https://doi.org/10.1021/jacs.8b05188}}.

\bibitem{Schmidt2025arxiv}
J.~Schmidt, D.~H. Ryan, O.~Janka, J.~K\"osters, C.~L. Mueller, A.~Sapkota, R.~F.~S. Penacchio, T.~J. Slade, S.~L. Bud'ko, P.~C. Canfield, {Suppression of the valence transition in solution-grown single crystals of ${\mathrm{Eu}}_{2}{\mathrm{Pt}}_{6}{\mathrm{Al}}_{15}$}, Phys. Rev. Mater. 9 (2025) 093404.
\newblock \href {https://doi.org/https://doi.org/10.1103/23zp-wpyc} {\path{doi:https://doi.org/10.1103/23zp-wpyc}}.

\bibitem{CanfieldP.C.KongT.KaluarachchiU.S.2016}
N.~H. Jo, P.~C. Canfield, T.~Kong, U.~S. Kaluarachchi, Use of frit-disc crucibles for routine and exploratory solution growth of single crystalline samples, Philos. Mag. 96 (2016) 84--92.
\newblock \href {https://doi.org/https://doi.org/10.1080/14786435.2015.1122248} {\path{doi:https://doi.org/10.1080/14786435.2015.1122248}}.

\bibitem{Canfield2020}
P.~C. Canfield, \href{https://doi.org/10.1088/1361-6633/ab514b}{New materials physics}, Rep. Prog. Phys. 83~(1) (2019) 016501.
\newline\urlprefix\url{https://doi.org/10.1088/1361-6633/ab514b}

\bibitem{Newbury2014}
D.~E. Newbury, N.~W.~M. Ritchie, {Rigorous quantitative elemental microanalysis by scanning electron microscopy/energy dispersive x-ray spectrometry (SEM/EDS) with spectrum processing by NIST DTSA-II}, Proc. SPIE 9236 (2014) 90--106.
\newblock \href {https://doi.org/https://doi.org/10.1117/12.2065842} {\path{doi:https://doi.org/10.1117/12.2065842}}.

\bibitem{Dreele2014}
R.~B. {Von Dreele}, {Small-angle scattering data analysis in GSAS-II}, J. Appl. Cryst. 47 (2014) 1784--1789.

\bibitem{Stoe2014}
{{STOE \& Cie GmbH}}, {X-Area (version 1.70)} (2014).

\bibitem{Fisher1962}
M.~E. Fisher, Relation between the specific heat and susceptibility of an antiferromagnet, Philos. Mag. 7 (1962) 1731--1743.
\newblock \href {https://doi.org/https://doi.org/10.1080/14786436208213705} {\path{doi:https://doi.org/10.1080/14786436208213705}}.

\bibitem{data_EuPt6Al16}
J.~Schmidt, O.~Janka, J.~K{\"o}sters, S.~L. Bud’ko, P.~C. Canfeld, {Dataset - Growth, discovery and characterization of single crystalline Eu$_{0.8}$Pt$_6$Al$_{16.4}$}, Zenodo (2025).
\newblock \href {https://doi.org/https://doi.org/10.5281/zenodo.17382556} {\path{doi:https://doi.org/10.5281/zenodo.17382556}}.

\bibitem{SI}
{See Supplemental Material at [insert link here] for the .cif file containing the information of the single-crystal refinements}.

\end{thebibliography}
\end{document}